\documentclass[12pt]{article}
\usepackage{sectsty}
\allsectionsfont{\sffamily}
\newcommand{\5}{$_5$}

\usepackage[nosort]{cite}

\usepackage[pdftex]{graphicx}
\usepackage[font={scriptsize,singlespacing}]{caption}
\usepackage{epstopdf}

\usepackage{amsmath}
\usepackage{amssymb}
\usepackage{subfigure}
\usepackage{hyperref}
\usepackage{url}
\usepackage{xcolor}
\usepackage{color}
\definecolor{amaranth}{rgb}{0.9, 0.17, 0.31}
\definecolor{purple(munsell)}{rgb}{0.62, 0.0, 0.77}
\definecolor{americanrose}{rgb}{1.0, 0.01, 0.24}
\definecolor{palatinateblue}{rgb}{0.15, 0.23, 0.89}
\definecolor{royalblue(web)}{rgb}{0.25, 0.41, 0.88}
\definecolor{hanpurple}{rgb}{0.32, 0.09, 0.98}
\definecolor{beaublue}{rgb}{0.74, 0.83, 0.9}
\definecolor{carminered}{rgb}{1.0, 0.0, 0.22}
\definecolor{brightpink}{rgb}{1.0, 0.0, 0.5}
\hypersetup{ linktoc=all,
    colorlinks, linkcolor={palatinateblue},
    citecolor={brightpink}, urlcolor={purple(munsell)}
}

\def\sideremark#1{\ifvmode\leavevmode\fi\vadjust{\vbox to0pt{\vss% the remark
 \hbox to 0pt{\hskip\hsize\hskip1em%                          will appear only
 \vbox{\hsize2cm\tiny\raggedright\pretolerance10000%          on the side
 \noindent #1\hfill}\hss}\vbox to8pt{\vfil}\vss}}}%
\newcommand{\bo}{\raise-1mm\hbox{\Large$\Box$}}

\newcommand{\be}{\begin{equation}}
\newcommand{\ee}{\end{equation}}
\newcommand{\bea}{\begin{eqnarray}}
\newcommand{\eea}{\end{eqnarray}}
\newcommand{\eqn}[1]{(\ref{#1})}

\begin{document}
\thispagestyle{empty}
\begin{center}

\null \vskip-1truecm \vskip2truecm

{\LARGE{\bf \textsf{Are Black Holes Springy?\\}}}

\vskip1truecm
\textbf{\textsf{Michael R.R. Good}}\\
{\footnotesize\textsf{Department of Physics, Nazarbayev University, \\53 Kabanbay Batyr Ave., 
Astana, Republic of Kazakhstan}\\
{\tt Email: michael.good@nu.edu.kz}}\\

\vskip0.4truecm
\textbf{\textsf{Yen Chin Ong}}\\
{\footnotesize \textsf{Nordic Institute for Theoretical Physics, \\ KTH Royal Institute of Technology \& Stockholm University, \\ Roslagstullsbacken 23,
SE-106 91 Stockholm, Sweden}\\
{\tt Email: yenchin.ong@nordita.org}}\\

\end{center}
\vskip1truecm \centerline{\textsf{ABSTRACT}} \baselineskip=15pt

\medskip
A $(3+1)$-dimensional asymptotically flat Kerr black hole angular speed $\Omega_+$ can be used to define an effective spring constant, $k=m\Omega_+^2$. Its maximum value is the Schwarzschild surface gravity, $k = \kappa $, which rapidly weakens as the black hole spins down and the temperature increases.  The Hawking temperature is expressed in terms of the spring constant: $2\pi T = \kappa - k$.  Hooke's law, in the extremal limit, provides the force $F = 1/4$, which is consistent with the conjecture of maximum force in general relativity.  

%{\color{blue}We also introduce an isoelastic process, namely fixing the ratio $k/\kappa$, and find no phase transition in the corresponding black hole thermodynamics. }

%\setcounter{tocdepth}{3}
\vskip0.4truecm
\hrule

\section{Introduction}

It has been a little more than 50 years since the Kerr solution, which describes a rotating black hole in general relativity, was discovered \cite{Kerr}. [For good reviews of the Kerr solution, see, e.g., \cite{Visser, Teukolsky}]. Nevertheless, there is still room left to either discover new things, or at least better understand some properties about this remarkable solution; for recent examples, see \cite{AL} in which the changing behavior of the Weyl curvature in the black hole exterior is found; and \cite{Jan} in which the properties of Killing vectors for the extremal Kerr metric is investigated in detail.

In the familiar Boyer-Lindquist coordinates, the Kerr geometry  [in the usual units $\hbar = c = G = k_B = 1$] is described by the following metric:
\begin{equation}
g[\text{Kerr}]=-\frac{\Delta}{\Sigma} \left(dt - a\sin^2\theta d\phi\right)^2 + \frac{\sin^2\theta}{\Sigma} \left[(r^2+a^2)d\phi - a dt\right]^2 + \frac{\Sigma}{\Delta}dr^2 + \Sigma d\theta^2,
\end{equation}
where 
\be
\Delta := r^2 - 2mr + a^2; ~~\Sigma:=r^2 + a^2\cos^2\theta,
\ee
and $a:=j/m$ is the angular momentum per unit mass of the black hole. Note that $a$ has the same dimension [of length] as the ADM mass $m$ in Planck units. 
The Kerr black hole has an outer event horizon at $r_+=m+\sqrt{m^2-a^2}$, and an inner Cauchy horizon at $r_-=m-\sqrt{m^2-a^2}$.

As is well-known, once we fix the mass and angular momentum for a Kerr black hole,
it will rotate with angular velocity 
\be \label{Omega}
\Omega_+ = \frac{a}{2mr_+} = \frac{a}{r_+^2 + a^2} \ee
at the outer event horizon. Note that this quantity is a constant for any isolated black hole at the classical level.

Once quantum effects are taken into account, Kerr black holes radiate with Hawking temperature given by the somewhat complicated expression
\be T = \frac{r_+ - r_-}{4\pi (r_+^2 + a^2)} = \frac{\sqrt{m^2 - a^2}}{4\pi m (m + \sqrt{m^2 - a^2})}. \ee 
The surface gravity \cite{Wald:1984rg}, $\kappa_{\textrm{Kerr}} = 2\pi T$, of the Kerr solution is given by, % \cite{Carroll:2004st} \cite{Fabbri:2005mw},
\be \label{kerrSG} \kappa_{\textrm{Kerr}}^2 = \frac{m^2 - a^2}{4 m^2 r_+^2}.\ee
Note that for a fixed mass $m$, a black hole with less surface area evaporates with lower temperature \cite{Bekenstein:1973ur} \cite{Hawking:1974sw}.  Since angular momentum reduces black hole surface area, it also serves to lower the temperature of the Kerr black hole, relative to an equal mass Schwarzschild black hole. [A similar role is of course played by the electric -- as well as the putative magnetic -- charges of the Reissner-Nordstr\"om black hole.]

According to the No-Hair Theorem, black holes in general relativity are exceptionally simple objects classified by only 3 parameters: mass, charge and angular momentum. It is therefore tempting to consider black holes as simple objects of study in quantum gravity. Of course, in the framework of quantum field theory, harmonic oscillators play a very important role\footnote{So much so that according to Sidney Coleman, ``\emph{the career of a young theoretical physicist consists of treating the harmonic oscillator in ever-increasing levels of abstraction}''.}. It is perhaps not surprising that black hole ring-down [the quasi-normal mode spectrum] has attracted a lot of attention; perhaps if we learn more about how black holes behave as a [classical] oscillator, we could make some progress in the still elusive realm of quantum gravity \cite{Berti}.   It has been conjectured that the added noise that necessarily accompanies a quantum measurement has a lower limit, which, using the quantum harmonic oscillator frequency applied to the quasi-normal mode frequency of black holes, gives the Hawking temperature \cite{Kiefer:2004ma}. It should be pointed out that quasi-normal modes are \emph{classical} and so it is somewhat of a mystery why this should have anything to say about the Hawking temperature, which is of course \emph{quantum} in nature. Regardless of whether this particular line of thought proves to be fruitful, it is an interesting question to ask whether black holes exhibit other properties that are analogous to harmonic oscillators.  Perhaps springs and black holes, as central objects of study in both quantum theory and general relativity respectively, may have more in common than is known.

In this work, we study the $(3+1)$-dimensional asymptotically flat Kerr black hole, which arguably is the \emph{most important black hole solution} -- as it is a very good approximation to the real, astrophysical black holes in our universe. We find that its angular speed can be used to define an effective spring constant. Although this only amounts to replacing the variables $(m,a)$ in the various equations with their functions $(k,\kappa)$, where $\kappa$ is the surface gravity for a Schwarzschild black hole with the same mass $m$, there are virtues in doing so. Indeed, not only do some equations, such as the expression for Hawking temperature, become clean and simple, this parametrization also allows us to calculate the rotational energy of the black hole easily [at least for small rotation regime]. We hope that more physical insights -- quantum or otherwise -- could be gained by introducing $k$ into the game.

\section{A Springy Kerr Black Hole}

Let us now see how the definition of $k$ is motivated. 
Using Eq.(\ref{Omega}) and inverting to obtain the angular momentum in terms of angular velocity, 
\be \label{Jinverted} j = \frac{4 m^3 \Omega_+}{1+4 m^2 \Omega_+^2}, \ee
as well as the maximum limit for the angular velocity of the event horizon [obtained from the fact that an extremal Kerr black hole has $m=a$], 
\be\label{limitofAV}\Omega_+ \leq \frac{1}{2m},\ee
the surface gravity of Eq.~\eqn{kerrSG}, after some simple algebra manipulations, can be expressed as
%\be \label{SGunfac}\kappa_{\textrm{Kerr}}^2 = \frac{1}{16m^2} - \frac{\Omega_+^2}{2} + m^2 \Omega_+^4.\ee
%Therefore,
%\be \label{SGrO} \kappa^2 = \frac{1}{4 r_+^2} - \Omega_+^2.\ee
%The negative contribution is readily seen but as a temperature, one obtains an infinite series,
%\be T = \frac{\kappa}{2\pi} = \frac{1}{4\pi r_+} - \frac{r_+ \Omega_+^2}{2\pi} + \mathcal{O}(\Omega_+^4), \ee
%in order to find the leading order term in angular velocity.  This can be avoided by 
%Expressing the surface gravity Eq. \eqn{SGrO} in terms of mass and angular velocity at the horizon,
\be \label{SGfac} \kappa_{\textrm{Kerr}}^2 = \left(\frac{1}{4 m} - m \Omega_+^2\right)^2. \ee
This form for the Kerr surface gravity suggests the introduction of a ``spring constant'' which serves to embody the rotational effect that diminishes the strength of the Kerr surface gravity, $\kappa_{\textrm{Kerr}}$, relative to the Schwarzschild surface gravity, $\kappa$:
\be \label{SGk} \kappa_{\textrm{Kerr}}^2 = \left(\kappa - k\right)^2. \ee
%This expression makes it simple to see that if there are two equal mass black holes, one spinning and the other is not, the angular momentum serves to lower the surface gravity.
In this context, the suggestion is that the spring constant, via simple harmonic motion, has thermodynamic relevance to the rotational surface gravity via the simple relation that the surface gravity is $2\pi$ times the Hawking temperature.

Indeed, the temperature of the Kerr black hole can now be expressed simply as

\be \label{mostsimple} \boxed{T = \frac{1}{2\pi}\left(\kappa - k\right)}, \ee 
%\be T = \frac{1}{8\pi m} - \frac{m}{2\pi} \Omega_+^2.\ee
%Reinstating units gives,
%\be T = \frac{\hbar c^3}{8 \pi G m k_B} - \frac{\hbar G }{2\pi c^3 k_B}m\Omega_+^2.\ee
where $\kappa$ is the Schwarzschild surface gravity, $\kappa = (4m)^{-1}$ [\emph{not} the Kerr surface gravity!], and $k$ is the harmonic spring constant\footnote{As we recall from elementary physics, for a mass attached to the end of a spring, the spring constant is $k=m\omega^2$, where $\omega=2\pi f$, in which $f$ is the frequency of the oscillatory motion. Our claim is that it may be helpful to ``identify'' this small omega with the capital omega of a rotating black hole.}, $k := m \Omega_+^2$. As a utilitarian use of parametrization alone, there is no new physics in Eq.\eqn{mostsimple}.  As such it is simply a convenience to separate the Schwarzschild surface gravity from the Kerr surface gravity because the former is sourced by the mass alone.  However, the source of rotation immediately finds itself thermodynamically characterized naturally by the black hole spring constant.  With interpretation, we consider $k$ to be a measure of the system stiffness of the rotating black hole, whose relationship to angular momentum is demonstrated in Section (\ref{section:stiffness}).  The spring constant for the rotating black hole exhibits the cooling effect due to spin.    The simplicity of this negative contribution suggests the comparison of radiating rotating black holes with simple harmonic oscillators, made in a somewhat similar vein as that done with rotating black holes and Newtonian rotating liquid drops \cite{Smarr:1972kt}. Reinstating units reveals the quantum nature of the spring constant contribution to temperature through the presence of $\hbar$, 
\be T = \frac{\hbar c^3}{8 \pi G m k_B} - \frac{\hbar G }{2\pi c^3 k_B}m\Omega_+^2. \ee
For perspective, a solar mass black hole rotating at 95\% the maximum rate has magnitude order $k\approx 10^{40}\; \textrm{N/m}$ compared to a spring in a ball point pen with constant of order about $k\approx 10^2 \; \textrm{N/m}$.   %, not the Kerr surface gravity of Eqs.~\eqn{kerrSG}, \eqn{SGfac}, and \eqn{SGk}. 

\section{Entropy of a Kerr Black Hole}
Appropriate to the conjecture that the Kerr black hole is dual to a $c_L = c_R = 12 j$ 2D CFT at $(T_L,T_R)$ for every value of $m$ and $j$ \cite{Castro:2010fd}, the number of states can be counted via
the Cardy formula \cite{Cardy} for microstate degeneracy,
\be S_{\text{micro}} = \frac{\pi^2}{3} (c_L T_L + c_R T_R). \ee
The ``temperatures''\footnote{These quantities do not have the same dimension as the usual temperature.} $T_{RL}$ are given by
\be T_L := \frac{m^2}{2\pi j}; \quad T_R := \frac{\sqrt{m^4 - j^2}}{2\pi j}. \ee
In terms of the spring constant $k$, we can re-write these expressions in an intriguing way:
\be T_{RL} =\frac{1}{4 \pi  \sqrt{k_*}}\mp \frac{\sqrt{k_*}}{4 \pi },\ee
where $k_* := k/\kappa$, and $0\leqslant k_* \leqslant 1$. Note that $T_R = T / \Omega_+ $.  The entropy of the Kerr black hole is then, assuming the central charges $c_L = c_R = 12 j$, and using Eq.~\eqn{Jinverted} and Eq.~\eqn{limitofAV}, 
\be S = \frac{\pi}{4 \kappa^2} (1+ k_*)^{-1}. \ee
This means that the surface area of the horizon is
\be \label{Areaspring} \boxed{A = \frac{\pi}{\kappa^2 + \kappa k}},\ee
which of course agrees with the well-known area formula obtained by simple integration of the Kerr metric:
\be \label{area} A = 4\pi(r_+^2 + a^2).\ee
This is
consistent with the fact that for two equal mass black holes, the one spinning more slowly has more area.  Higher entropy in this physical picture corresponds to a decrease in $k$ or less stiffness.

%%%%%%%%%%%%%%%%%%%%%%%%%%%%%%%%%%%%%%%%%%%%%%%%%%%%%%%%%%%%%%%%%%%%%%%%%%%%%%%%%%

\section{Simple Harmonic Potential Energy From Rotational Energy in the Limit of Slow Rotation}

We recall that for the Kerr-Newman family, it is useful to define the so-called ``irreducible mass'', $m_{\textrm{irr}}$. Essentially, it is the energy that \emph{cannot} be extracted from a black hole via any classical process. A Schwarzschild black hole simply has $m_{\textrm{irr}}$ equal to its ADM mass. However, for Kerr black holes, the rotational energy can be extracted from say, the Penrose process \cite{Penrose}, so that its irreducible mass is smaller than its ADM mass.

The irreducible mass for the Kerr black hole is given explicitly as \cite{Christodoulou:1970wf}
\be
m_{\textrm{irr}}^2 = \frac{A}{16\pi}.
\ee
By the second law of black hole mechanics, this quantity is non-decreasing in time. Note also that the irreducible mass saturates the Riemannian Penrose inequality.

In terms of $k$ and $\kappa$, we have
\be
m_{\textrm{irr}}^2 = \frac{1}{16(\kappa^2 + \kappa k)}.
\ee
It can be checked that the quantity, 
\be \label{ksubstar} k_* :=\frac{k}{\kappa} = \frac{j^2}{4 m_{\textrm{irr}}^4},\ee
ranges from zero to one, where the energy expression [often called the \emph{Christodoulou-Ruffini Mass Formula}]
\be  m^2 = m_{\textrm{irr}}^2 + k_* m_{\textrm{irr}}^2 \ee
holds.  

The smallest irreducible mass, $m_{\textrm{irr}}=m/\sqrt{2}$, when $k = \kappa$, or equivalently when $j=m^2$, gives the maximum energy that can be extracted from the black hole:
\be\frac{m-m_{\textrm{irr}}}{m} = 1 - \frac{1}{\sqrt{2}} \approx 0.29. \ee
The maximal spring constant is reduced to zero via reversible Penrose processes, consistent with the familiar removal of $29\%$ of the initial extremal black hole energy.  The rotational energy of the black hole, $E_r = m- m_{irr}$, expressed in parameters $(m,j)$, is
\be \label{rotmj} E_r = m- \frac{1}{\sqrt{2}} \sqrt{\sqrt{m^4 - j^2}+m^2} = \frac{j^2}{8 m^3} + \mathcal O(j^4).\ee
In terms of the parameters $(k_*,\kappa)$, this becomes
\be \label{Er} E_r = \frac{1}{4 \kappa }-\frac{1}{4\kappa \sqrt{ 1+k_*}}  = \frac{k_*}{8\kappa} + \mathcal O(k_*^2).\ee
In terms of $(k,r_+)$, one has instead
\be E_r = \frac{r_+}{2}  \left(\frac{1}{1-2 k r_+}-\frac{1}{\sqrt{1-2 k r_+}}\right). \ee
Consistent with virtually any oscillatory motion that has small amplitude, one finds for small $r_+$,
\be E_r = \frac{1}{2} k r_+^2 + \mathcal O(r_+^3).\ee
Also, for small $k$, the first order term remains the simple harmonic potential energy,
\be E_r = \frac{1}{2} k r_+^2 + \mathcal O(k^2).\ee
Comparing with Eq.(\ref{rotmj}), we conclude that for slow rotation, $j \ll m^2$, we have
\be \frac{1}{2} k r_+^2 = \frac{j^2}{8 m^3}, \ee
and furthermore we see that, from Eq.(\ref{Er}),
\be \frac{k_*}{8\kappa} = \frac{j^2}{8 m^3}.\ee
Therefore, with the spring constant introduced, simple harmonic potential energy provides an easy and rather good approximation for the rotational energy of slow spinning black holes.  

%\section{Minimal Noise Temperature}

%The uncertainty relation places a weaker limit on the harmonic oscillator's position measurement, compared to that of the added noise that occurs when an amplified signal undergoes processing.  Assuming a linear amplifier which is phase-insensitive, and unitarity, one defines the noise number $A$ and the measured number of quanta, the gain number $G$ via 

%\be A \geq \frac{1}{2} | 1 \mp G^{-1} | . \ee

%Noise is added to the signal for high gain $G>1$.  For vanishingly small temperature, the quantum harmonic oscillator energy per mode relates the noise number to the frequency:

%\be \frac{1}{2} + \frac{1}{e^{\omega / T}-1} = \frac{1}{2} \coth \frac{\omega}{2 T} \equiv A + \frac{1}{2}.\ee

%At absolute zero, the noise number $A = 0$.  The lower limit for the noise temperature $T = T_n$ is then expressed in terms of the gain  

%\be T_n \geq \omega \ln^{-1} \left( \frac{3 \mp G^{-1}}{1\mp G^{-1}}\right).\ee

%For a highly damped signal, such as the case of the Schwarzschild black hole quasi-normal frequency, the limit of infinite gain, $G \rightarrow \infty$, must be taken,

%\be T_n = \frac{\omega}{\ln 3}. \ee

%This is the lower limit on the noise temperature.  For fixed angular momentum $\ell$ the quasi-normal modes has a countable infinite number whose real part is

%\be \omega_{QNM} = \frac{\kappa}{2\pi} \ln 3. \ee

%Setting the input frequency of the oscillator to quasi-normal mode, $\omega = \omega_{QNM}$ yields the minimum noise temperature for a quantum measurement of the highly damped quasi-normal modes:

%\be T_n = \frac{\kappa}{2\pi}. \ee

\section{Maximal Tension in the Extremal Limit}
Consideration of a rotating black hole as ``Hookean'' or linear-elastic [this means we \emph{define} the force associated with spring constant $k$ by $F=kx$], begs the question of the limiting nature at maximum system stiffness.  The maximum spring constant value, angular velocity value and angular momentum value are, respectively, 
\be \label{maxvalues} k^{\textrm{max}} = \kappa, \quad \Omega_{+}^{\textrm{max}} = \frac{1}{2m}, \quad j^{\textrm{max}}=m^2.\ee
The maximally stiff system, by Hooke's law, has limiting maximum force, 
\be F = k x \Rightarrow \frac{1}{4m}\left(\lim_{j \to m^2}{r_+}\right) = \frac{1}{4m}{m} = \frac{1}{4}, \ee
consistent with the conjecture\footnote{The conjecture states that the maximum tension or force between two bodies cannot exceed $F=c^4/{4G} \approx 3.25 \times 10^{43} ~\text{N}$. While the combination $c^4/G$ gives the right dimension, the factor $1/4$ is non-trivial, but seems to hold in various examples \cite{Gibbons:2002iv}. Note also the absence of $\hbar$.} of Gibbons \cite{Gibbons:2002iv}, Schiller \cite{chris}, and Barrow and Gibbons \cite{Barrow:2014cga}.  Here we see an indication on how this maximum force is realized in the context of rotating black holes.  Simple harmonic motion is often considered as a one-dimensional projection of uniform circular motion.  In this result the projection follows the straightforward assignment of $\Omega_+$ as the angular velocity around the outer event horizon radius, $r_+$, to simple harmonic motion with angular frequency $\Omega_+$ and amplitude $r_+$, where $k = m \Omega_+^2$.  As it is common for materials to resist being compressed beyond a certain minimum size, or stretched beyond a maximum size, without some permanent deformation or change of state, one would expect black hole linear-elasticity to be bounded.  Hooke's law in the high spring constant limit for the case of an extremely fast spinning black hole suggests relevance of black hole elasticity to the maximum force of general relativity.  

Since Hooke's law is a first order linear approximation to the real response of physical springs, it might be expected to wildly fail far before maximum or extreme system stiffness, in the way that many materials so noticeably deviate from Hooke's law well before elastic limits are reached.  However, Hooke's law is an accurate approximation for solid bodies when deformations are small, and in the fast spinning Kerr case, because the elasticity is so high we might be allowed to interpret the black hole deformation to be minimal, accompanying a minimal area [see Eq.~\eqn{Areaspring}].  Considering this, we find that Hooke's law results in the same limit for maximal force in general relativity with zero cosmological constant in $3+1$ dimensions as conjectured in \cite{Gibbons:2002iv, chris, Barrow:2014cga}.

%\section*{Thermodynamics and Lifetime at Constant Elasticity Ratio}

\section{System Stiffness as Black Hole Spins Down} \label{section:stiffness}
\begin{figure}[htbp]
\centering
\includegraphics[width=7.5cm,keepaspectratio]{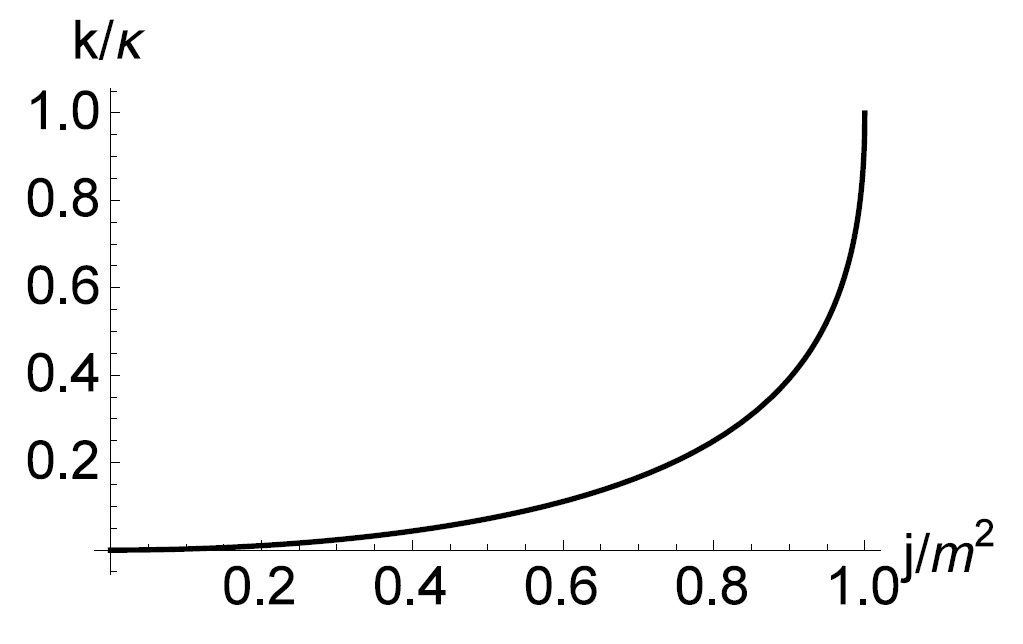}
\caption{\label{fig:1} The spring constant dependence on angular momentum demonstrates an initial rapid approach to zero from maximum angular momentum. }
%\end{minipage}
\end{figure}

Consider the angular momentum's relationship to the spring constant, both of which are taken to be normalized with respect to their maximum value, $k_* := k/\kappa$, and $j_*:= j/m^2$. We have: 
\be \label{star} j_* = \frac{2\sqrt{k_*}}{1+k_*},\quad k_*=\frac{2}{j_*^2}-\frac{2 \sqrt{1-j_*^2}}{j_*^2}-1.\ee
Here we have chosen the negative sign in the inversion of $k_*$. In the limit of zero angular momentum where $j_* \rightarrow 0$, the elasticity must also $k_* \rightarrow 0$.  This behavior is consistent with the choice of the negative sign.  Otherwise, $k_* \rightarrow \infty$ in the limit of $j_* \rightarrow 0$, which grossly violates the maximum value $k = \kappa$.  The spin-stiffness relationship is conveyed in Figure \ref{fig:1}. We see a sharp decrease in elasticity accompanying an initial decrease from maximum angular momentum, e.g. $50\%$ of the stiffness is lost while $94\%$ of the angular momentum still remains.   

Many materials have an elastic modulus that decreases with the temperature \cite{PhysRevB.2.3952}. Often, matter becomes less stiff, and more easy to stretch and deform as temperature increases.  The black hole spring constant softens to zero as the temperature rises. 
%The maximally stiff system, by Hooke's law, has limiting force, $F = 1/4$, consistent with \cite{Gibbons:2002iv}.

Using Eq.\eqn{area} for the horizon area, Eq.\eqn{mostsimple} for the Hawking temperature and Stefan-Boltzmann law for the total power emitted [assuming perfect black-body emission], $P = \sigma A T^4$, gives
\be P = \frac{(\kappa - k)^4}{960 \pi  \kappa  (\kappa +k)}. \ee
To calculate the lifetime of the black hole in this case, it is fully tractable if one assumes any \emph{isoelastic} process that allows the black hole to maintain a constant elasticity ratio $k_* = k/\kappa$. Since 
\begin{equation}
k_*=\frac{m\Omega_+^2}{1/4m}=\frac{a^2}{r_+^2},
\end{equation}
by Eq.(\ref{Omega}), we have
\begin{equation}
k_*=\frac{\left(\frac{a}{m}\right)^2}{2-\left(\frac{a}{m}\right)^2+2\sqrt{1-\left(\frac{a}{m}\right)^2}}.
\end{equation}
Therefore we see that an isoelastic process is equivalent to fixing the ratio $a/m$, or equivalently $j_*=j/m^2$, as can be seen from Eq.(\ref{star}). We do not speculate on the probability for an astrophysical rotating black hole being subject to such a process [indeed the ``standard'' scenario of Hawking evaporation will steadily\footnote{See however, the appendix of \cite{YJS}, in which it is argued that black holes may spin \emph{up} initially due to the backreaction from Hawking radiation.} decrease the ratio $a/m$ to 0 in a finite, albeit long, time \cite{page}].  The power emitted in terms of a time independent $k_*$ is
\be P = -\frac{d}{d t}m(t) = \frac{\left(k_*-1\right)^4}{15360 \pi  \left(k_*+1\right)} m(t)^{-2}.\ee
The mass  $m(t)$ can be solved upon imposing initial condition $m(0)=m_0$; it is
\be m(t) = m_0^3-\frac{\left(k_*-1\right)^4 t}{5120 \pi  \left(k_*+1\right)}, \ee
which gives the lifetime until $m(t)\rightarrow 0$ as
\be t = \frac{5120 \pi  (k_*+1) m_0^3}{(k_*-1)^4}. \ee
The effect of high stiffness [$k_* \approx 1$] serves to greatly lengthen the lifespan.  

\begin{figure}[!h]
\centering
\mbox{\subfigure{\includegraphics[width=3.0in]{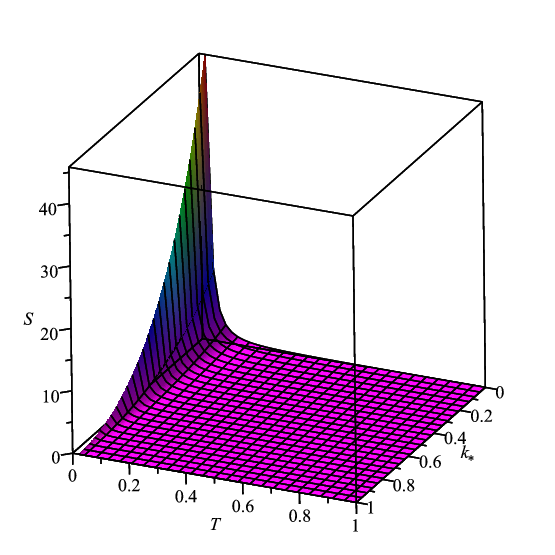}}\quad
\subfigure{\includegraphics[width=3.0in]{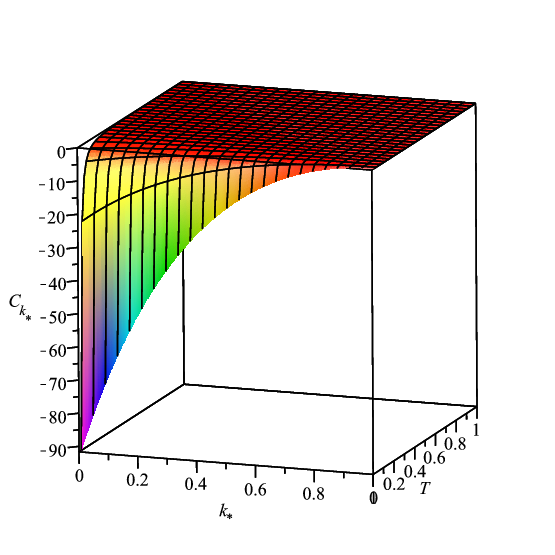} }}
\caption{\textbf{Left:}  The entropy $S$ as a function of the Hawking temperature $T$ and the normalized spring constant $k_*=k/\kappa$. \textbf{Right:} The specific heat in an isoelastic process is always negative; there is no phase transition. As is obvious from Eq.\eqn{siso} and Eq.\eqn{ciso}, $C_{k_*}=-2 S$. \label{STK}}
\end{figure}

If one considers the entropy of the Kerr black hole as a function of the Hawking temperature $T$ and the normalized spring constant $k_*$, we have the rather succinct expression:
\begin{equation} \label{siso}
S(T,k_*)= \frac{(k_*-1)^2 }{16 \pi T^2(k_*+1)}.
%S(T,k_*)=\frac{1}{8\pi} \left(\frac{(k_*-1)^2 }{2T^2(k_*+1)}\right).
\end{equation}
This is plotted on the left diagram in Fig.(\ref{STK}).
This parametrization allows one to compare specific heats of black hole thermodynamics, which is more simple than working with $S(T,j_*)$, 
\begin{equation}
S(T,j_*)=\frac{1}{8\pi(j_*^2T^2)}\left[(1-j_*^2) - \sqrt{1-j_*^6 + 3j_*^4 - 3j_*^2}\right],
\end{equation}
despite the fact that fixing $k_*$ is physically equivalent to fixing $j_*$.  We now see that in an isoelastic process, the specific heat $C_{k_*}$ is given by
\begin{equation} \label{ciso}
C_{k_*}:=T\left.\frac{\partial S}{\partial T}\right\vert_{k_*} = -\frac{(k_*-1)^2}{8\pi T^2(k_*+1)} < 0.
\end{equation}
Since the denominator of this expression is always non-negative, and only vanishes in the extremal case $T=0$, there is no phase transition in an isoelastic process -- the specific heat is \emph{always negative} [right diagram in Fig.(\ref{STK})]. 

We remark that the phrase ``phase transition'' is used conventionally in the black hole context whenever there is a change of sign in the specific heat. From example, Ressiner-Nordstr\"om black holes may under-go phase transition \emph{twice} in some regimes of their evolutions \cite{kn:HW}. Phase transition in this sense does \emph{not} necessarily mean that the black hole geometry changes drastically [cf. phase transition of liquid water to solid ice]. The issue of black hole instability is somewhat tricky and specific heat does not contain all the subtleties. In particular this should not be confused with true phase transition such as Hawking-Page phase transition \cite{HP} from Schwarzschild-AdS black hole to thermal AdS. [In that case, Hawking-Page phase transition occurs for black hole with horizon $r_h < L$, whereas the naive ``phase transition'' due to the change of specific heat occurs at $r_h=L/\sqrt{3}$.] 
 
Comparing to the equilibrium thermodynamics of the Kerr black hole which is governed by the Gibbs free energy \cite{Altamirano:2014tva}, the specific heat at constant pressure\footnote{In this context, it is proposed that pressure can be defined as $P=-\Lambda/8\pi$, where $\Lambda < 0$ is the cosmological constant in an asymptotically AdS spacetime. The first law of black hole mechanics is generalized to $dM = T dS + \Phi dQ + \Omega_+ dJ + V dP$, where $V$ is the ``thermodynamic volume'' \cite{pdv}. Here in asymptotically flat case $P\equiv 0$, but $C_P$ still makes sense.} 
\be C_P = \frac{\pi(1-k_*)}{2\kappa^2(3k_*(2+k_*)-1)} \ee
has two branches.  The branch where $C_P$ is positive requires the bound
 \be \frac{2 \sqrt{3}}{3} - 1 \approx  0.154 < k_* < 1, \ee
implies in a canonical ensemble, more springy black holes are thermodynamically preferred over less springy black holes.  This behavior should also be compared to $C_j$ and $C_a$, the specific heats obtained by holding $j$ and $a$ constant, respectively. Both of these exhibit phase transitions as shown by Davies \cite{davies1,davies2,davies3}-- at the values $j^2/m^4 = 2\sqrt{3}-3 \approx 0.464$ and $j^2/m^4 =(\sqrt{5}-1)/2 \approx 0.618$, respectively\footnote{One of Davies' papers \cite{davies1} contains a mistake which led him to conclude that phase transition $C_j$ occurs at $j^2/m^4 =(\sqrt{5}-1)/2$; he did however obtain the correct answer years earlier \cite{davies2, davies3}. Nevertheless one could show that for $C_a$, there \emph{is} a phase transition at  $j^2/m^4 =(\sqrt{5}-1)/2$. See the blog of John Baez \cite{baez} for more discussions.}.

We remark in passing that the value $(\sqrt{5}-1)/2$ is of course the inverse of the golden section $\varphi= (\sqrt{5}+1)/2 \approx 1.618$, which satisfies the unique property: $\varphi^{-1} = \varphi -1$. This occurrence of the golden section in black hole physics is intriguing and perplexing [though it might be just a coincidence]. It is interesting to recall from elementary classical mechanics exercise that if we attach two identical springs with spring constant $k$ and two identical masses $m$ in a series, with the free end attached to a wall, then the normal mode frequencies are precisely
\begin{equation}
\omega_{\pm} = \sqrt{\frac{k}{m}}\frac{\sqrt{5} \pm 1}{2}.
\end{equation} 
That is, the golden section [and its inverse] appears as the value relating the frequency of two springs connected in series to the frequency of a single spring.
In a spring-black hole context, it is tempting to think that this is more than a coincidence, and that the springy property of Kerr black hole is somehow manifest in the specific heat $C_a$. However, we are not able to justify this supposition.

%As the black hole spins down, the increase in surface area is seen by smaller value of $k$ in the entropy
%\be S = \frac{\pi}{4\kappa^2 + 4\kappa k} \ee

\section{Discussion}

In this work we introduced an effective spring constant $k$ for a $(3+1)$-dimensional asymptotically flat Kerr black hole, and studied various properties of the black hole in terms of the surface gravity $\kappa$ for Schwarzschild black hole of the same mass and $k$. This simplifies various expressions and allows a simple calculation of the rotational energy of the black hole in the slow rotation limit. Notably the Hawking temperature takes the simple form
\be
T=\frac{1}{2\pi}\left(\kappa-k\right),
\ee
which makes it evident that for any fixed mass, rotation cools down the black hole, and correspondingly in the springy language, the stiffness decreases as temperature increases\footnote{In Reissner-Nordstr\"om spacetime, an analogous calculation can be carried out, and one obtains $T=(\kappa-k_q)/(2\pi)$, where $k_q := q^4/(4mr_+^4) \equiv \kappa \Phi^4$ in which $q$ is the electrical charge carried by the black hole. Here $\Phi$ is an electric potential in an appropriate gauge where the potential is zero at infinity.  Unlike the uncharged rotating case however, we are not aware of a ``natural'' elasticity interpretation for $k_q$.}.  Hooke's Law once applied to Kerr black holes, is also compatible with the maximal force $F=1/4$ proposed for general relativity. 

%In principle, using the variables $(k,\kappa)$ instead of the usual $(m,a)$ can have some caculational advantages in at least some contexts, e.g. Hawking temperature, surface gravity, . 

%For example, from the area formula Eq.(\ref{Areaspring}), using the fact that $0 \leq k \leq \kappa=1/(4m)$, it is almost immediately evident that 
%\be
%\frac{\pi}{16(\kappa^2 + \kappa k)} \leq m^2,
%\ee
%that is, the Riemannian Penrose Inequality holds for Kerr black hole, with equality if and only if $k=0$. Of course for this simple example, it is not so difficult to see this result using the variables $(m,a)$ either. \edz{We can place non-trivial ($k$, $\kappa$) simplified examples of the Kerr black hole here.}

Motivated by the idea that harmonic oscillatory considerations of black holes may provide some insights into quantum gravity, let us end the discussion by examining an interpretation for the ground state [``zero-point''] energy of a Kerr black hole. Recall that the Hawking temperature of a black hole vanishes in the extreme limit. It is thus often argued that  ``naturally''\footnote{This is modulo some subtleties about the [classical, as well as stringy] instability of extremal black holes \cite{0601001,0905.1180,1110.2006,1206.6598,1208.1437,1307.6800}. In the context of AdS/CFT correspondence, it is actually also problematic from the field theory perspective to have a stable extremal black hole, for then its entropy remains finite at zero temperature; see Footnote 14 of \cite{Hartnoll}.} we should interpret extremal black holes as ground states of the corresponding quantum theory. The ground state energy of an extremal black hole [at zero absolute temperature] using the same association used to find $2 \pi T = \kappa - k$, but now considering a \emph{quantum} harmonic oscillator, sets $\omega\rightarrow \Omega_+$ in the zero temperature limit.  The maximum angular velocity of Eq.~\eqn{maxvalues} is used and the ground state energy corresponding to an extremal black hole [$\hbar=1$] is
\be E_0 = \frac{\hbar \omega}{2} = \frac{\Omega_+^{\textrm{max}}}{2} = \frac{1}{4 m} = \kappa, \ee
which is surprisingly just the surface gravity of a Schwarzschild black hole of the same mass. This is consistent with interpreting extremal rotating black holes as the ground states of the corresponding quantum theory.  Upon reinstating the units, the ground state energy of a Kerr black hole is
\be E_0 = \frac{\hbar c^3}{4 G} \frac{1}{m} = \frac{\hbar}{m c} \frac{c^4}{4G} \approx 10.6\; m^{-1} \; \textrm{Joules}. \ee
The reduced Compton wavelength of the black hole multiplied by the maximum force gives this energy.

We shall end by repeating the question posted in the title: are black holes really springy, or are all these just coincidences in the case of $(3+1)$-dimensional asymptotically flat Kerr geometry, among its ``many properties which have endowed the Kerr metric with an aura of the miraculous'' \cite{chandrasekhar}? The latter seems to be more likely. For example, it does not work for Kerr black holes in AdS. As another example,
one could consider an asymptotically anti-de Sitter black string \cite{FS, CZ, LZ}. Its surface gravity is\footnote{In our unit, a Schwarzschild black hole has temperature $1/8\pi m$, i.e., the temperature [and thus also the surface gravity] has dimension of inverse length. However we note that here, due to the non-compactness of black string, $M$ is defined as the ADM mass \textit{density} along the string, so $M$ is dimensionless. This gives the right dimension to the temperature.}
\begin{equation} \label{1}
\kappa[a]=3\alpha \left(\frac{M}{2}\right)^{1/3} \sqrt{1-\frac{\alpha^2a^2}{2}}\left(1-\frac{3\alpha^2a^2}{2}\right)^{-\frac{1}{6}},
\end{equation}
where  $\alpha^2=-\frac{\Lambda}{3}=1/L^2 > 0$ is the inverse of the square of AdS length scale. The parameter $a$ controls the rotation rate of the black string, its range is such that $0 \leqslant \alpha a \leqslant 1$.
In the static case we have,
\begin{equation} \label{2}
\kappa := \kappa[a=0] = 3\alpha \left(\frac{M}{2}\right)^{1/3}.
\end{equation}

For the black string, we have instead, from Eq.(\ref{1}) and Eq.(\ref{2}), 
\begin{equation}
k:=\kappa-\kappa[a] = 3\alpha \left(\frac{M}{2}\right)^{1/3} \left[1-\sqrt{1-\frac{\alpha^2a^2}{2}}\left(1-\frac{3\alpha^2a^2}{2}\right)^{-\frac{1}{6}}\right].
\end{equation}
The angular speed of the black string is\cite{LZ} 
\begin{equation}
\Omega_{+}=\frac{a\alpha^2}{\sqrt{1-\frac{a^2\alpha^2}{2}}}.
\end{equation}
For small $a$, it can be shown that
\begin{equation}
k\approx \frac{M^{1/3}}{2^{1/3}} \left(\frac{3a^2\alpha}{8}\right)\Omega_+^2. 
\end{equation}
This is in fact ``natural'' -- since $M$ is dimensionless, the only possible parameter of the right dimension available is $a$, so that $k/\Omega_+^2$ has to be proportional to $\alpha^{n} a^{n+1}$, $n \in \Bbb{N}$. This behavior is markedly different from the asymptotically flat Kerr case in which $k = m \Omega_+^2 $. 

In addition, since one of the [probably necessary]  reasons why this works for asymptotically flat Kerr black hole in 3+1 dimensions is due to the fact that simple harmonic motion can be treated as a one-dimensional projection of uniform circular motion, higher dimensional cases would likely \emph{not} work. There are also complications that arise from the fact that in higher dimensions, black holes can rotate with respect to multiple axes of rotation. Furthermore it is well-known that in dimensions 6 and above, there is no extremal limit for Kerr black holes. It might still be interesting to investigate what happens to the higher dimensional generalizations of the maximum force conjecture \cite{Gibbons:2002iv} if one naively defines $k$ in a similar way and applies Hooke's Law. 
Nevertheless, perhaps there is a deeper reason why the asymptotically flat Kerr black hole in 3+1 dimensions admits such an effective spring constant description.  We hope that this question will contribute to the ongoing effort of further understanding the remarkable Kerr solution.

\section*{Acknowledgment}
The authors thank the Institute of Advanced Study of Nanyang Technological University in Singapore, at which this project was first discussed.

%%%%%%%%%%%%%%%%%%%%%%%%%%%%%%%%%%%%%%%%%%%%%%%%%%%%%%%%%%%%%%%%

\end{document}